\documentclass[preprint2]{aastex6}
\begin{document}
\title{Stacking Spectra in Protoplanetary Disks: Detecting Intensity Profiles from Hidden Molecular Lines in HD~163296}

\author{Hsi-Wei Yen\altaffilmark{1,2}, Patrick M. Koch\altaffilmark{1}, Hauyu Baobab Liu\altaffilmark{2}, Evaria Puspitaningrum\altaffilmark{3}, Naomi Hirano\altaffilmark{1}, Chin-Fei Lee\altaffilmark{1}, Shigehisa Takakuwa\altaffilmark{1,4} }

\altaffiltext{1}{Academia Sinica Institute of Astronomy and Astrophysics, P.O. Box 23-141, Taipei 10617, Taiwan} 
\altaffiltext{2}{European Southern Observatory (ESO), Karl-Schwarzschild-Str. 2, D-85748 Garching, Germany; hyen@eso.org}
\altaffiltext{3}{Department of Astronomy, Faculty of Mathematics and Natural Sciences, Institut Teknologi Bandung, Jl. Ganesha 10, Bandung 40132 Indonesia}
\altaffiltext{4}{Department of Physics and Astronomy, Graduate School of Science and Engineering, Kagoshima University, 1-21-35 Korimoto, Kagoshima, Kagoshima 890-0065, Japan}

\begin{abstract}
We introduce a new stacking method in Keplerian disks that 
(1) enhances signal-to-noise ratios (S/N) of detected molecular lines and 
(2) that makes visible otherwise undetectable weak lines. 
Our technique takes advantage of the Keplerian rotational velocity pattern.  
It aligns spectra according to their different centroid velocities at their different positions in a disk and stacks them. 
After aligning, the signals are accumulated in a narrower velocity range as compared to the original line width without alignment.
Moreover, originally correlated noise becomes de-correlated.
Stacked and aligned spectra, thus, have a higher S/N. 
We apply our method to ALMA archival data of DCN (3--2), DCO$^+$ (3--2), N$_2$D$^+$ (3--2), and H$_2$CO (3$_{0,3}$--2$_{0,2}$), (3$_{2,2}$--2$_{2,1}$), and (3$_{2,1}$--2$_{2,0}$) in the protoplanetary disk around HD~163296.  
As a result, (1) the S/N of the originally detected DCN (3--2), DCO$^+$ (3--2), and H$_2$CO (3$_{0,3}$--2$_{0,2}$) and N$_2$D$^+$ (3--2) lines are boosted by a factor of $\gtrsim$4--5 at their spectral peaks, implying one order of magnitude shorter integration times to reach the original S/N; 
and (2) the previously undetectable spectra of the H$_2$CO (3$_{2,2}$--2$_{2,1}$) and (3$_{2,1}$--2$_{2,0}$) lines are materialized at more than 3$\sigma$.  
These dramatically enhanced S/N allow us to measure intensity distributions in all lines with high significance. 
The principle of our method can not only be applied to Keplerian disks but also to any systems with ordered kinematic patterns.
\end{abstract}

\keywords{ISM: individual objects (HD~163296) --- line: profiles --- protoplanetary disks}

\section{Introduction}
Studying the physical and chemical structures of protoplanetary disks is essential to understand the environment of planet formation (e.g., Dutrey et al.~2014). 
With the Atacama Large Millimeter/submillimeter Array (ALMA), protoplanetary disks can be observed 
with resolutions $<$0\farcs1 
(e.g., ALMA partnership et al.~2015). 
Although molecular lines can be simultaneously observed with ALMA, 
their images are typically generated at resolutions of a few times 0\farcs1, lower than continuum images (e.g., ALMA partnership et al.~2015; Qi et al.~2015), because the sensitivity of line observations is limited by their narrow line widths (on the order of 10 km s$^{-1}$).
Therefore, intensity distributions of molecular lines
cannot be directly compared with (sub-)millimeter continuum data.
Besides, even with ALMA, there are faint molecular lines that remain difficult to detect.
In order to advance our understanding of physical and chemical structures in protoplanetary disks, 
it is, thus, important to develop techniques to enhance signal-to-noise ratios (S/N) of molecular-line data
and reveal their intensity distributions at higher angular resolutions. 

An optimal filtering technique is introduced in Dutrey et al.~(2007) to enhance S/N of integrated intensities of molecular-line emission in Keplerian disks.
Their technique, however, relies on a best-fit line profile that needs to be derived first from model fitting. This is then adopted to weigh velocity channels for integration.
Here, 
we introduce a new method to enhance the detection of molecular lines while conserving the total integrated flux with minimum assumptions on line profiles.
Our method takes advantage of the ordered and symmetric Keplerian rotational velocity pattern. 
In a Keplerian disk, spectra at different positions have different centroid velocities. 
Different from directly stacking spectra from different positions in a disk -- which is not coherently adding signals --
our method first aligns these spectra based on their expected Keplerian velocities, and then azimuthally and/or radially stacks them. 
With this alignment, 
the total integrated flux of the stacked spectrum remains unchanged, but is redistributed into a narrower velocity range, as compared to stacking without prior alignment.
As a result, the peak intensity increases, and thus, the S/N is boosted. 
Stacking techniques are also applied in extragalactic studies.
In galaxy samples, spectra of individual galaxies can be redshift-aligned and then stacked to enhance a line detection yielding sample-averaged properties (e.g., Delhaize et al.~2013).
In a single resolved galaxy, molecular-line spectra from different positions can be aligned using observed mean velocities of H{\sc i} at the same positions as a proxy. They can then be stacked to measure gas properties at different positions (e.g., Cald\'u-Primo et al.~2013; Higdon et al.~2015).
In this paper, we focus on Keplerian disks, and we demonstrate our method using ALMA archival data of 
the protoplanetary disk around HD~163296. 

\begin{figure}
\epsscale{1}
\plotone{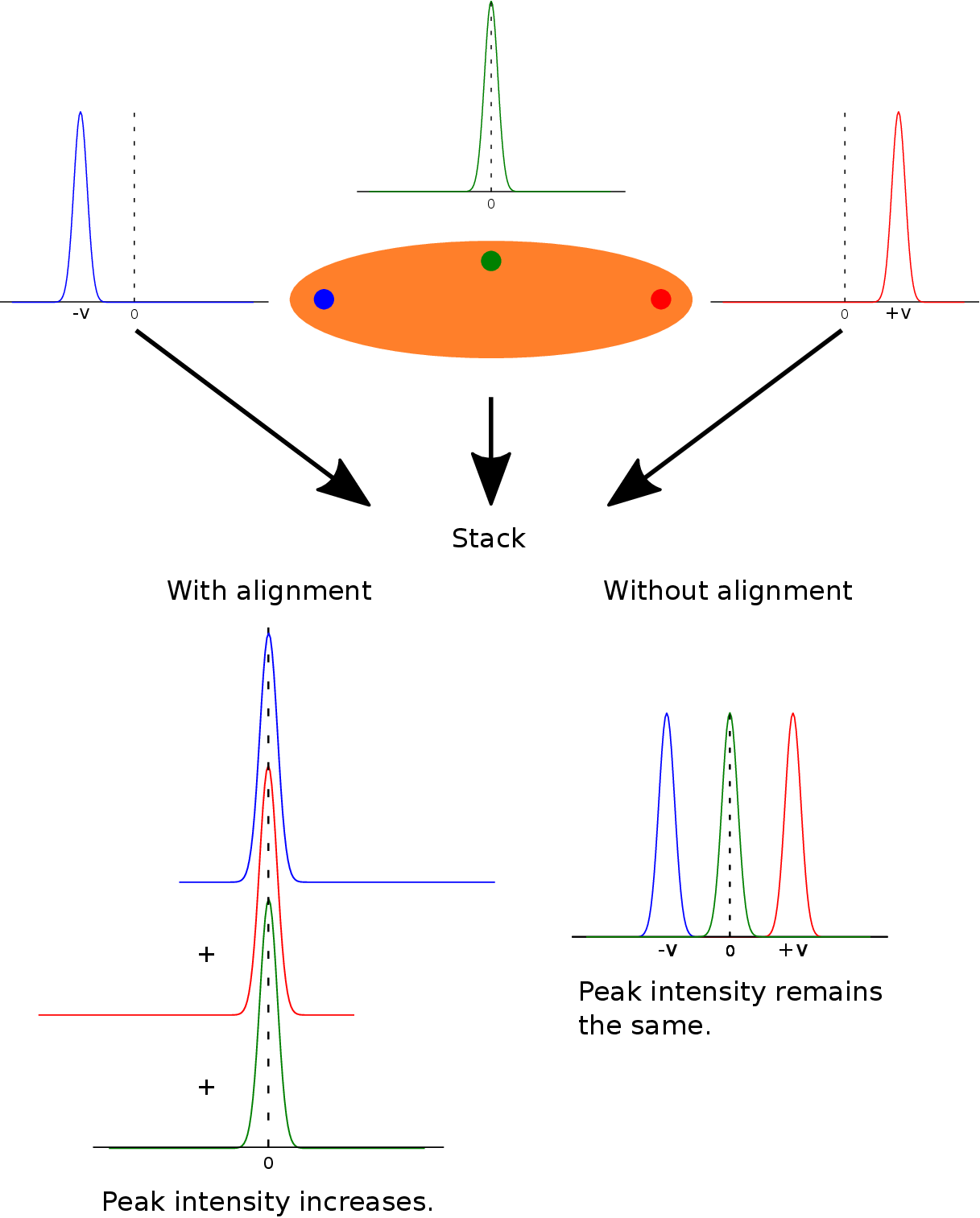}
\caption{Schematic illustration of stacking three spectra having different centroid velocities (color curves) at different positions (color dots) in a Keplerian disk (orange oval). When aligning the spectra, the signals are shifted to the same velocity, and the line width becomes narrower. As a result, the peak intensity and, thus, the S/N increases after stacking. Without alignment, the signals remain at different velocities,
do not coherently add after stacking, and the S/N is not enhanced.}
\label{align}
\end{figure}

\section{New Stacking Method}
Our method makes use of the Keplerian rotational velocity pattern, which is derived with the inclination ($\tbond i$), the position angle of the disk major axis ($\tbond PA$) and stellar mass ($\tbond M_*$). 
$M_*$, $i$, and $PA$ can be measured from continuum and molecular-line images with sufficient S/N. 
We first correct for the plane-of-sky projection and compute the expected Keplerian velocity at every pixel in the disk. 
Assuming that the disk is geometrically thin, the projection correction is
\begin{eqnarray}
x &=& \Delta\alpha\sin PA + \Delta\delta\cos PA, \nonumber \\
y &=& \Delta\delta\sin PA -\Delta\alpha\cos PA,
\end{eqnarray}
where $x$ and $y$ are the positional offsets along the major and minor axes, and $\Delta\alpha$ and $\Delta\delta$ are the RA and Dec offsets with respect to the central star. 
The de-projected radius ($\tbond r$) in the disk plane is computed as
\begin{equation}
r = \sqrt{x^2 + \left(\frac{y}{\cos i}\right)^2}, 
\end{equation}
and the projected Keplerian velocity ($\tbond V_{\rm k}$) at the position ($\Delta\alpha$, $\Delta\delta$) as
\begin{equation}\label{vkep}
V_{\rm k} = \sqrt{\frac{GM_*}{r}}\cdot \cos\theta\cdot\sin i,
\end{equation}
where $G$ is the gravitational constant and $\cos\theta = x/r$.
We first align spectra ($\tbond I_V$) from different positions ($\Delta\alpha_j$, $\Delta\delta_j$) by shifting them by their $V_{{\rm k},j}$($\Delta\alpha_j$, $\Delta\delta_j$). We then stack them,  i.e., add coherently in velocity,  as
\begin{equation}
\sum_{j=1}^N I_V(\Delta\alpha_j, \Delta\delta_j, V-V_{{\rm k},j}),
\end{equation}
where $V$ is the relative velocity with respect to the systemic velocity of the disk. 
$N$ is the number of pixels within the radial and azimuthal ranges adopted for stacking.
Figure \ref{align} illustrates the stacking process.
Additionally, since the Keplerian rotational velocity pattern is mirror-symmetric with respect to the disk minor axis, 
spectra from one side of the minor axis are first reversed with respect to their $V_{\rm k}$ before being stacked with spectra from the other side$\footnotemark[1]$. 
Alignment is not possible for small $r$ when $V_{\rm k} \propto r^{-0.5}$ is beyond the velocity band.
Such pixels can still be stacked, but the alignment cannot be applied. 

\footnotetext[1]{For blended hyperfine components, our method can still be applied using their major components to align their overall line profiles. However, the S/N of minor hyperfine components is not fully enhanced because of their inaccurate alignment due to slight frequency differences and their line profiles which are not symmetric with respect to $V_{\rm k}$ of their major components. Nevertheless, the overall S/N can be enhanced. Non-blended components can simply be treated as individual lines.}

In our approach, we coherently stack (i.e., add after aligning) spectra from different positions in a Keplerian disk.
In the following we derive approximate formulae for the achieved S/N gain in this process. We explain the two effects that make this possible, namely (1) narrower line widths and (2) the de-correlation of initially correlated noise.
In order to be explicit, we illustrate our calculations for azimuthally averaged spectra. 

\subsection{Narrower Line Widths}\label{Valign}
Without any alignment, 
the line width of an azimuthally averaged spectrum ($\Delta V_{\rm na}$) at a radius $r$ in a Keplerian disk is approximately
\begin{equation}
\Delta V_{\rm na}(r) \sim 2 \times V_{\rm k} (r,\theta) \sim 2 \sqrt{\frac{GM_*}{r}}\sin i, 
\end{equation}
where $\theta=0$ is adopted because $V_{\rm k}$ reaches its maximum along the major axis.
The factor of two results from $V_{\rm k}$ having opposite signs (blue- and red-shifted) with respect to the systemic velocity on the two sides of the disk.
The line width of an azimuthally averaged spectrum after alignment in velocity ($\Delta V_{\rm a}$) is related to its intrinsic line width ($\Delta V_{\rm int}$) and line broadening due to beam convolution ($\Delta V_{\rm conv}$) over positions with different $V_{\rm k}$. 
$\Delta V_{\rm int}$ is a combination of thermal and turbulent line widths together with the integration along the line of the sight passing through different scale heights, where the rotational velocities are different. 
Observations of a sample of protoplanetary disks show that the typical 1$\sigma$ line width of $\Delta V_{\rm int}$ is  
\begin{equation}
\Delta V_{\rm int} \sim(0.1\mbox{--}0.3) \times (\frac{r}{100\ {\rm AU}})^q\ {\rm km\ s^{-1}},
\end{equation}
where $q$ ranges from $-0.3$ to $-0.1$ (e.g., Pi{\'e}tu et al.~2007). 
$\Delta V_{\rm conv}$ can be described as
\begin{equation}\label{vcon}
\Delta V_{\rm conv} \sim \frac{A_{\rm b}}{r} \times \frac{\partial^2}{\partial r \partial \theta} V_{\rm k} \sim \frac{A_{\rm b}\sqrt{GM_*}}{2r^{2.5}}\sin\theta\sin i,
\end{equation}
where $A_{\rm b}$ is the area of the synthesized beam. $\Delta V_{\rm a}$ is then estimated as
\begin{equation}
\Delta V_{\rm a} \sim 4\Delta V_{\rm int} + \Delta V_{\rm conv}, \label{eq_delta_va}
\end{equation}
where  the factor of four results from considering the line width covering most of the total integrated flux, i.e., the full-width-zero-intensity (FWZI) line width, which is approximately 4$\Delta V_{\rm int}$.

\subsection{De-correlation of Correlated Noise in Interferometric Images}
Interferometric images are generated by Fourier-transforming visibilities. 
Thus, in an interferometric image, pixels within a synthesized beam area are not independent but are correlated, 
while pixels in different velocity channels are independent from each other. 
Therefore, when averaging over an area ($A_{\rm ave}$) without alignment in velocity, the number of independent pixels is approximately $A_{\rm ave}/A_{\rm b}$. 
On the contrary, 
when computing an average spectrum after velocity alignment, 
averaging is performed over pixels that originally were in different velocity channels.
As a result, the number of independent pixels increases. 
Generally,  
the number of independent pixels increases when the difference in $V_{\rm k}$ between nearby pixels is larger than the velocity channel width $dv$. 
The spatial scale $A_{\rm dep}$ where the difference in $V_{\rm k}$ becomes larger than $dv$ can be derived as
\begin{equation}\label{Adep}
A_{\rm dep} = \frac{rdv}{\frac{\partial^2}{\partial r \partial \theta}V_{\rm k}} \sim \frac{2r^{2.5}dv}{\sqrt{GM_*}\sin\theta\sin i},    
\end{equation}
and only pixels within $A_{\rm dep}$ are correlated after alignment.  
Hence, the number of independent pixels when averaging over $A_{\rm ave}$ after alignment becomes $A_{\rm ave}/A_{\rm dep}$. 
We note that $A_{\rm dep}$ has a limited range, 
\begin{equation}
A_{\rm pix} \leq A_{\rm dep} \leq A_{\rm b},  
\end{equation}
where $A_{\rm pix}$ is the area of one pixel. 
This is because the minimum scale in an image is one pixel, and in interferometric images pixels separated by more than the synthesized beam size are inherently not correlated.

\subsection{Approximate Formulae for S/N Enhancement after Stacking with Alignment}
Since the total integrated intensity over all velocity channels remains unchanged after alignment, 
the peak intensity of azimuthally averaged spectra after alignment in velocity increases approximately as $\Delta V_{\rm na} / \Delta V_{\rm a}$.
Noise, on the contrary, decreases as the square root of the number of independent measurements after averaging, 
i.e., by $\sqrt{A_{\rm b}/A_{\rm dep}}$. 
As a result, 
the S/N of azimuthally averaged spectra at the peak after alignment is boosted by 
\begin{equation}
\frac{\Delta V_{\rm na}}{\Delta V_{\rm a}} \times \sqrt{\frac{A_{\rm b}}{A_{\rm dep}}}, 
\end{equation}
while the S/N of the total integrated intensity or the mean intensity per channel of azimuthally averaged spectra is enhanced after alignment by
\begin{equation}
\sqrt{\frac{\Delta V_{\rm na}}{\Delta V_{\rm a}}} \times \sqrt{\frac{A_{\rm b}}{A_{\rm dep}}}, 
\end{equation}
where $\Delta V_{\rm na}$ and $\Delta V_{\rm a}$ are explicitly derived in Section \ref{Valign}.
The S/N enhancement in azimuthally averaged spectra after alignment ($R_{\rm sn}$) can then be estimated as
\begin{equation}\label{Rsn}
R_{\rm sn} \sim \sqrt{\frac{\Delta V_{\rm na}}{4\Delta V_{\rm int} + \Delta V_{\rm conv}}} \times \sqrt{\frac{A_b}{A_{\rm dep}}}.
\end{equation}
At outer radii, the velocity gradient ($\partial_r V_{\rm k}$) is small, $4\Delta V_{\rm int} \gg \Delta V_{\rm conv}$, and thus, 
\begin{equation}\label{Rsn1}
R_{\rm sn} \sim \sqrt{\frac{\Delta V_{\rm na}}{4\Delta V_{\rm int}}} \times \sqrt{\frac{A_b}{A_{\rm dep}}} \propto r^{-\frac{3+q}{2}}.
\end{equation}
At inner radii, where $\Delta V_{\rm conv} \gg 4\Delta V_{\rm int}$, 
\begin{equation}\label{Rsn2}
R_{\rm sn} \sim \sqrt{\frac{\Delta V_{\rm na}}{\Delta V_{\rm conv}}} \times \sqrt{\frac{A_b}{A_{\rm dep}}} \propto r^{-0.25}.
\end{equation}
Therefore, the S/N enhancement increases with smaller radii and flattens at larger radii.
The improvement is small when $V_{\rm k}$ is beyond the velocity band and when $r$ is small compared to the beam size.

The S/N enhancement further depends on the accuracy of $V_{\rm k}$ at individual positions. 
Therefore, the (unknown) scale heights of molecular distributions in a disk, uncertainties in stellar mass and disk orientation, and coarse velocity resolutions lead to more inaccurately aligned spectra, leaving the signal spread over a wider velocity range even after alignment.
Hence, these effects reduce the S/N of stacked spectra after alignment, and a maximum S/N is achieved when correct stellar mass and disk orientation are adopted. 
Consequently, this method can also be used to calibrate estimated stellar mass and disk orientation. 
Since the S/N enhancement is related to the ratio of line widths of spectra with and without alignment, 
this method is more effective for disks originally having wider line widths, i.e., disks with larger stellar masses and closer to edge-on.  
Stacking with alignment can be performed over any meaningful combination of radial and azimuthal ranges
as long as the chosen area leads to the desired S/N.
Hence, this method can also be used
to extract intensity profiles (when stacked azimuthally in radial bins) or azimuthally asymmetric distributions (when stacking azimuthally distinct sectors). 

\section{Demonstration Case: HD~163296}

HD~163296 is a Herbig Ae star at a distance of 122 pc (van den Ancker et al.~1998).
It is surrounded by a several hundred-AU disk showing a clear Keplerian rotation with a central stellar mass $\sim$2.5 $M_\sun$ (from CO, HCO$^+$, and their isotopes; Mannings \& Sargent 1997; Isella et al.~2007; Hughes et al.~2008; de Gregorio-Monsalvo et al.~2013; Mathews et al.~2013; Rosenfeld et al.~2013). 
Inclination and position angle are measured to be $\sim$45$\arcdeg$ and $\sim$130$\arcdeg$ (Isella et al.~2007).
Its three-dimensional structure is imaged in CO with ALMA (de Gregorio-Monsalvo et al.~2013; Rosenfeld et al.~2013). 
Observations in $^{13}$CO, DCO$^+$, and N$_2$H$^+$ have revealed the location of the CO snow line at $r \sim 90$ AU (Qi et al.~2011, 2015; Mathews et al.~2013). The ALMA DCO$^+$ and N$_2$H$^+$ images clearly show ring-like structures (Mathews et al.~2013; Qi et al.~2015). 
Several additional molecular lines toward HD~163296 were selected by ALMA but remain marginally or not at all detected. 
This disk is, thus,  an excellent proof-of-concept target.

The HD~163296 data analyzed here are retrieved from the ALMA archive (project code: 2013.1.01268.S). 
Observations were done with 31 to 33 antennas during the cycle-2 observing period on July 27--29, 2014. 
HD~163296 was observed for 3.6 hours in total. 
The pointing center was $\alpha$(J2000) = $17^{h}56^{m}21\fs28$, $\delta$(J2000) = $-21\arcdeg57\arcmin22\farcs4$. 
The correlator was configured in the Frequency Division Mode.
DCO$^+$ (3--2; 216.113 GHz), DCN (3--2; 217.239 GHz), N$_2$D$^+$ (3--2; 231.322 GHz), and H$_2$CO (3$_{0,3}$--2$_{0,2}$; 218.222 GHz), (3$_{2,2}$--2$_{2,1}$; 218.476 GHz), and (3$_{2,1}$--2$_{2,0}$; 218.76 GHz) were observed simultaneously with a spectral resolution of 61 kHz. 
Each spectral window had a bandwidth of 58.6 MHz. 
Calibration of the raw visibility data was performed with the standard reduction script for cycle-2 data using tasks in Common Astronomy Software Applications (CASA), and without self-calibration.
Calibrated visibilities were Fourier-transformed with natural weighting and CLEANed with the CASA task ``clean'' at a velocity resolution of 0.1 km s$^{-1}$. 
The angular resolutions of the images are $\sim$0\farcs5 $\times$ 0\farcs4. 
The noise levels are 2.4 mJy Beam$^{-1}$ in the DCO$^+$ and DCN, 2 mJy Beam$^{-1}$ in the H$_2$CO, and 2.9 mJy Beam$^{-1}$ in the N$_2$D$^+$ image.

\begin{figure*}
\epsscale{1}
\plotone{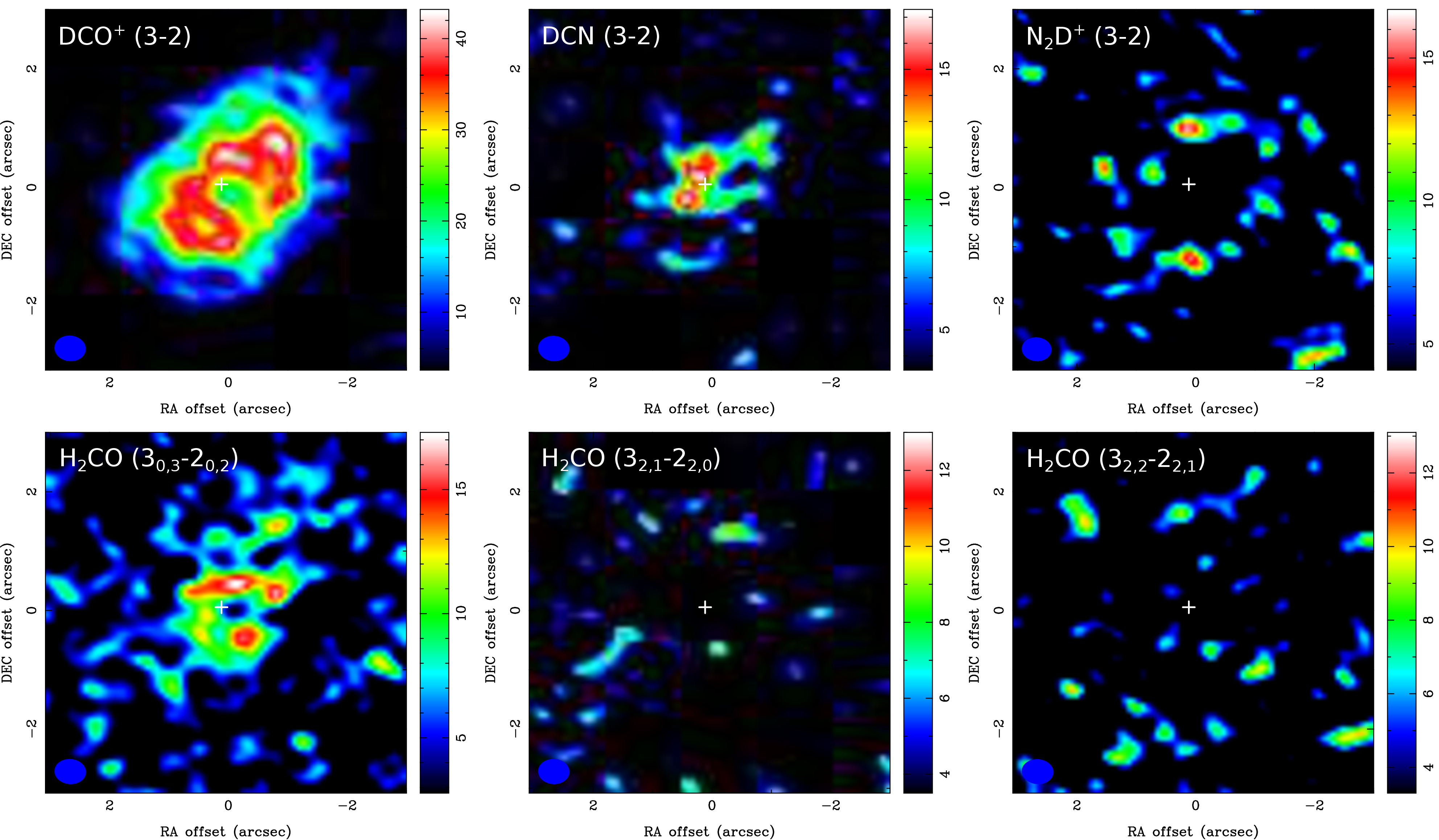}
\caption{Moment 0 maps in units of mJy Beam$^{-1}$ km s$^{-1}$ of DCO$^+$ (3--2), DCN (3--2), N$_2$D$^+$ (3--2), and H$_2$CO (3$_{0,3}$--2$_{0,2}$), (3$_{2,2}$--2$_{2,1}$), and (3$_{2,1}$--2$_{2,0}$) in HD~163296 from ALMA archival data. White crosses denote the stellar position, and blue filled ellipses present the beam sizes.}
\label{mom0}
\end{figure*}

\section{Results and Discussion}
\begin{figure*}
\epsscale{1}
\plotone{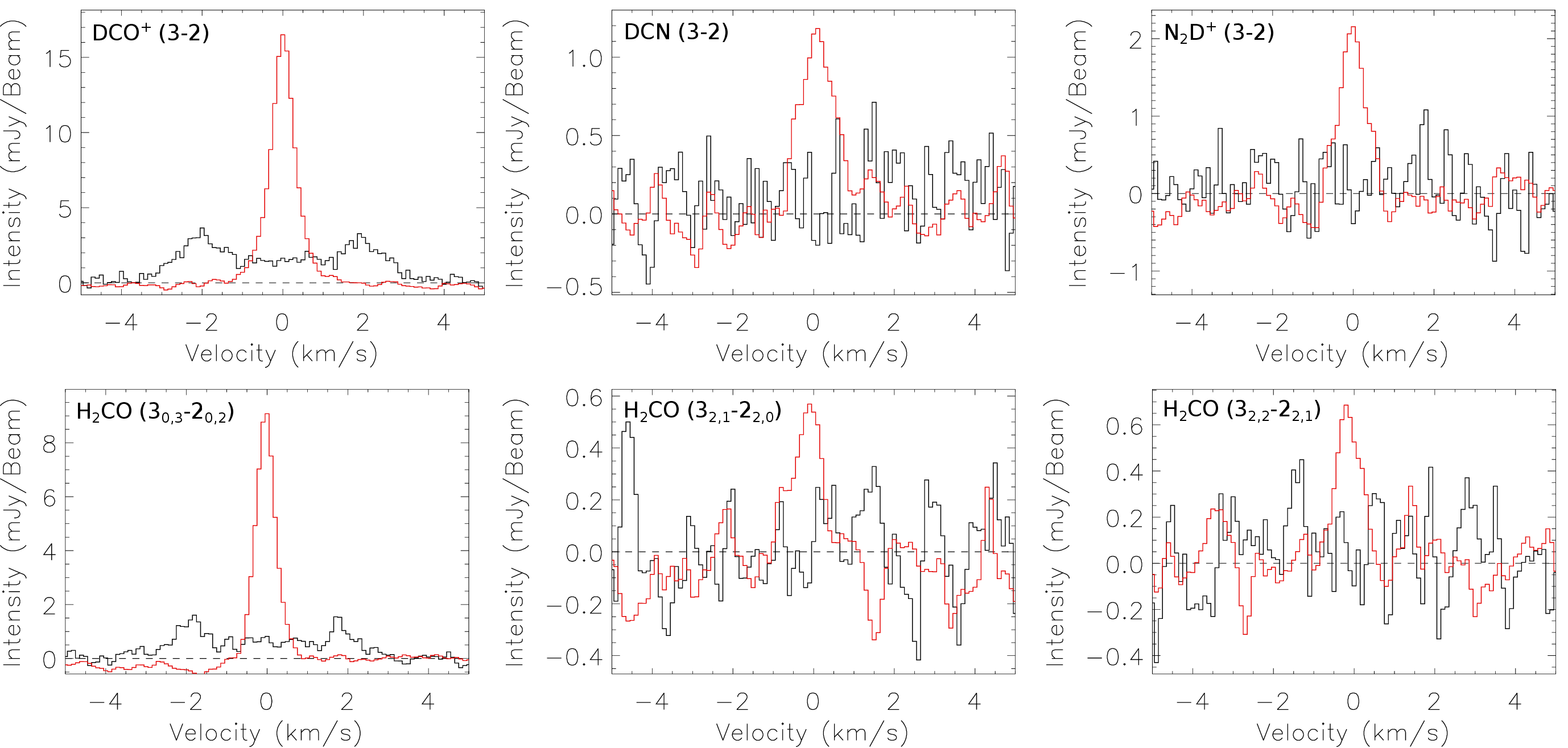}
\caption{Averaged (radial and over full azimuthal range from 0 to $2\pi$) spectra of DCO$^+$ (3--2), DCN (3--2), N$_2$D$^+$ (3--2), and H$_2$CO (3$_{0,3}$--2$_{0,2}$), (3$_{2,2}$--2$_{2,1}$), and (3$_{2,1}$--2$_{2,0}$) in HD~163296. Black histograms present the conventional averaged spectra from the original image cubes, and red ones are from our method that aligns the centroid velocities of the spectra according to their Keplerian velocities at different positions within the disk area before averaging.}
\label{spec}
\end{figure*}

Figure \ref{mom0} presents the total integrated intensity (moment 0) maps of the six molecular lines from archival data. 
DCO$^+$ (3--2) and H$_2$CO (3$_{0,3}$--2$_{0,2}$) are clearly detected at S/N $\gtrsim$10, showing ring-like structures. 
Such structures are also seen in ALMA DCO$^+$ (4--3) and N$_2$H$^+$ (3--2) observations of HD~163296 (Mathews et al.~2013; Qi et al.~2015), 
and H$_2$CO (3$_{1,2}$--2$_{1,1}$) and (4$_{1,4}$--3$_{1,3}$) are suggested showing ring-like structures from observations with the Submillimeter Array (Qi et al.~2013).
DCN (3--2) and N$_2$D$^+$ (3--2) are also detected in our moment 0 maps at S/N $\sim$5. 
DCN (3--2) is offset from the center and more compact than DCO$^+$ (3--2) and H$_2$CO (3$_{0,3}$--2$_{0,2}$). 
N$_2$D$^+$ (3--2) displays several clumpy components, hinting a ring-like structure. 
The H$_2$CO (3$_{2,2}$--2$_{2,1}$) and (3$_{2,1}$--2$_{2,0}$) maps show no clear detections. 
Figure \ref{spec} exhibits azimuthally averaged spectra without alignment in velocity (black histograms). 
The radial ranges for averaging are adopted to be $R < 300$ AU ($\sim$2\farcs5) for DCO$^+$ (3--2) and H$_2$CO (3$_{0,3}$--2$_{0,2}$),  (3$_{2,2}$--2$_{2,1}$) and (3$_{2,1}$--2$_{2,0}$), $R < 200$ AU ($\sim$1\farcs6) for DCN (3--2), and $100 < R < 300$ AU ($\sim$0\farcs8--2\farcs5) for N$_2$D$^+$ (3--2).
As we show below,  these radial ranges are the regions where the emission lines primarily originate from.
The spectra of DCO$^+$ (3--2) and H$_2$CO (3$_{0,3}$--2$_{0,2}$) are clearly detected at S/N $>$ 10.
They show double peaks, the characteristic line profile of Keplerian rotation. 
The spectra of DCN (3--2) and N$_2$D$^+$ (3--2) are seen at $<$4$\sigma$. 
H$_2$CO (3$_{2,2}$--2$_{2,1}$) and (3$_{2,1}$--2$_{2,0}$) are neither detected in their moment 0 maps, nor their spectra, nor their image cubes.

The red histograms in Figure \ref{spec} display azimuthally averaged spectra with alignment, following our method, 
where averaging is over the same radial and azimuthal ranges as above.
The black histograms show original spectra with a line width $\sim$6 km s$^{-1}$ for DCO$^+$ (3--2) and H$_2$CO (3$_{0,3}$--2$_{0,2}$).
After applying our method,  
the signals are accumulated in a much narrower velocity range of $\sim$1 km s$^{-1}$.
As a result, the peak intensities of the spectra increase by a factor $\sim$5, thus boosting their S/N. 
Fluxes integrated over the full velocity range of the new spectra of DCO$^+$ (3--2) and H$_2$CO (3$_{0,3}$--2$_{0,2}$) are measured to be 12.9$\pm$0.1 and 4.9$\pm$0.1 mJy Beam$^{-1}$ km s$^{-1}$, and are 
consistent with those from the original spectra (11.4$\pm$0.2 and 4.5$\pm$0.3 mJy Beam$^{-1}$ km s$^{-1}$). 
DCN (3--2) and N$_2$D$^+$ (3--2) are seen more clearly at $>$10$\sigma$ in the averaged spectra with alignment.
Their S/N are enhanced by a factor of $\gtrsim$4.
Strikingly, H$_2$CO (3$_{2,2}$--2$_{2,1}$) and (3$_{2,1}$--2$_{2,0}$) -- which are not detected at all in the moment 0 maps nor the spectra in the original data -- are now clearly visible.

\begin{figure*}
\epsscale{1}
\plotone{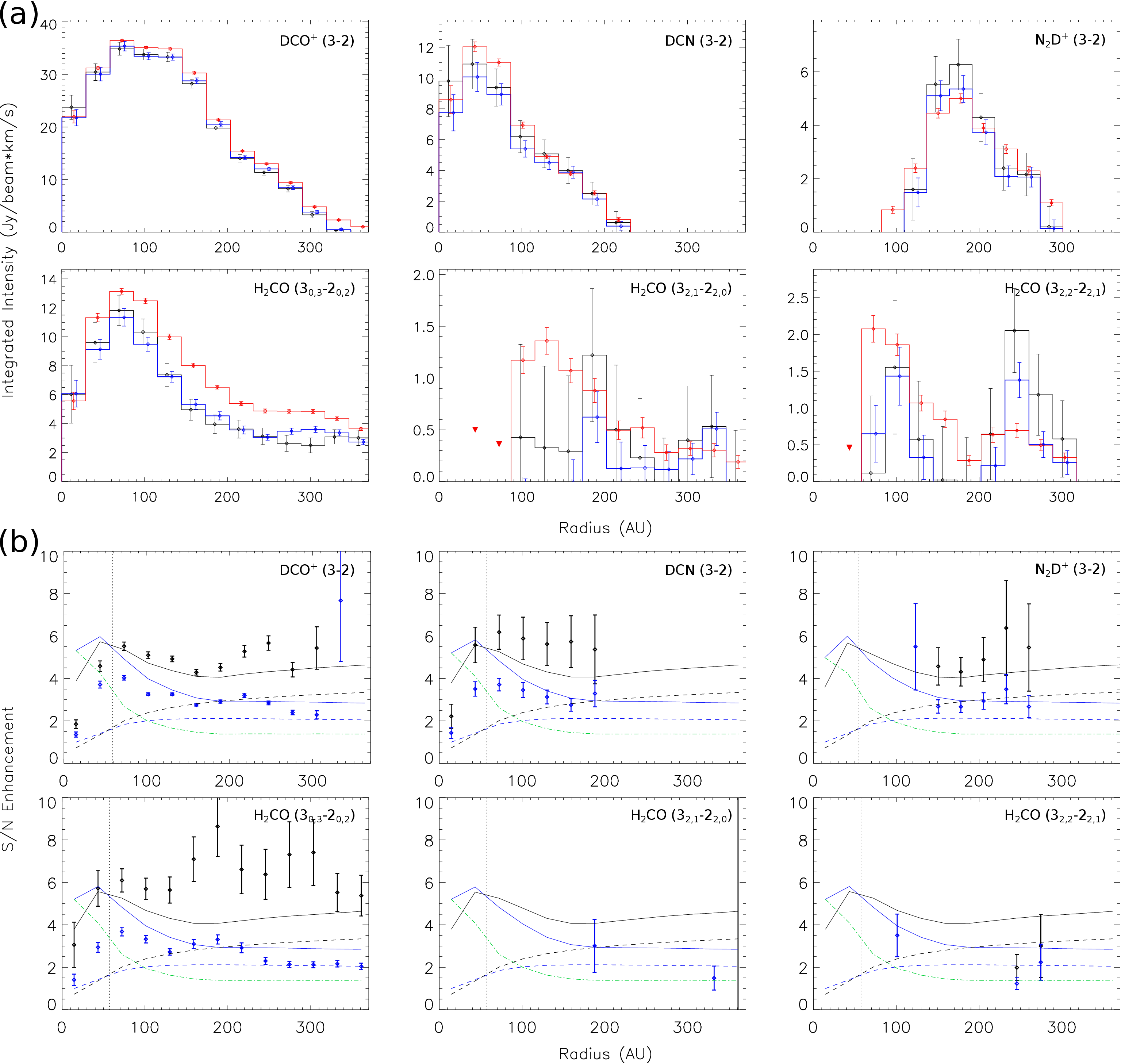}
\caption{(a) Radial intensity profiles of DCO$^+$ (3--2), DCN (3--2), N$_2$D$^+$ (3--2), and H$_2$CO (3$_{0,3}$--2$_{0,2}$), (3$_{2,2}$--2$_{2,1}$), and (3$_{2,1}$--2$_{2,0}$) in HD~163296. Black, blue, and red histograms present the profiles derived with (1) moment 0 maps, (2) averaged spectra without alignment, and (3) averaged spectra with alignment. Error bars correspond to 1$\sigma$ uncertainties. The profiles from all three methods are measured at the same radial bins. For clarity, black and blue data points are slightly off-set from the centers of the radial bins. (b) Radial profiles of 
S/N enhancements (data points) of measured intensity profiles from averaged spectra with alignment compared to method 
(1) using moment 0 maps (black) and method (2) with averaged spectra without alignment (blue). The enhancement is computed by dividing the S/N from method (3) by those from method (1) and (2). Error bars are computed by propagating 1$\sigma$ uncertainties. Solid lines present the expected enhancements from Equation \ref{Rsn}. Dashed lines show the enhancement due to narrower line withs after alignment (first term in Equation \ref{Rsn}), and green dashed-dotted lines show the enhancement due to noise de-correlation (second term in Equation \ref{Rsn}).}
\label{Ir}
\end{figure*}

In Figure \ref{Ir}a, we compare radial intensity profiles derived with three different methods: (1) from the original moment 0 maps, (2) from azimuthally averaged spectra without alignment, and (3) from azimuthally averaged spectra with alignment$\footnotemark[2]$. 
The profiles from all three methods have the same radial bins and are averaged over the full azimuthal range of 2$\pi$. 
Each radial bin is half of the synthesized beam size.
The intensity profiles extracted from the moment 0 maps are derived by computing mean values of pixels in each radial bin, 
with their uncertainties estimated as
\begin{equation}
\frac{\sqrt{N_0}\cdot \sigma \cdot dv}{\sqrt{A_{\rm ave} / A_{\rm b}}},
\end{equation} 
where $N_0$ is the number of velocity channels included to generate the moment 0 maps, $\sigma$ is the noise per velocity channel in the original image cubes, 
and $A_{\rm ave}$ corresponds to the area of each radial bin. 
For the intensity profiles derived from azimuthally averaged spectra with and without alignment, 
we first stack spectra only from pixels in the same radial bin with and without alignment, respectively. 
Then we compute the integrated intensity and divide it by the number of the pixels in the radial bin to derive a mean integrated intensity at each radius.
Their uncertainties are estimated as
\begin{equation}
\sigma_{\rm a} \cdot dv \cdot \sqrt{N}, 
\end{equation} 
where $\sigma_{\rm a}$ is the noise per channel in the averaged spectra and $N$ is the number of integrated velocity channels.
As explained in Section 2, $\sigma_{\rm a}$ and $N$ of the averaged spectra with alignment are smaller than those without alignment because of noise de-correlation and narrower line widths.

\footnotetext[2]{We also applied our alignment method to dirty maps to extract intensity profiles. The profiles are consistent with those from CLEANed maps within $\sim$5\%--10\% at radii $\lesssim$250 AU ($\sim$2\arcsec), which is the radius of the first strong negative side lobes at a $\sim$3\% level. This comparison shows that our intensity profiles are largely unaffected by imperfect deconvolution.}

Since the total integrated flux is conserved in all three methods, 
intensity profiles necessarily need to be consistent within uncertainties.  
This is, indeed, observed for DCO$^+$ (3--2) and DCN (3--2) which are detected at high significance in the original data. 
In contrast, 
the two weakest lines, H$_2$CO (3$_{2,2}$--2$_{2,1}$) and (3$_{2,1}$--2$_{2,0}$), seem to show differences.
This is because their intensities are comparable to their uncertainties when directly working with their moment 0 maps or their averaged spectra without alignment.  
These intensity profiles are, thus, highly uncertain.
Here, method (3) is clearly superior, greatly reducing uncertainties and being the only one of the three methods that is able to measure radial features. 

In Figure~\ref{Ir}b we compare gains in S/N for the intensity profiles from the three different methods.
Only data points at radial bins where all three methods show a detection at more than 3$\sigma$ are plotted. 
DCO$^+$ (3--2) and DCN (3--2) show ratios in the S/N of intensity profiles from averaged spectra with alignment that are $\sim$4--6 times higher than those from moment 0 maps, and they are $\sim$2--4 times higher than those from averaged spectra without alignment. 
Our results further show that only with the boost in S/N provided by our method of alignment,  
the intensity profiles of the weaker H$_2$CO (3$_{2,2}$--2$_{2,1}$) and (3$_{2,1}$--2$_{2,0}$) lines can be measured at high significance. 
Additionally, the integrated intensities computed from averaged spectra with alignment tend to be higher than those from method (1) and (2). 
This is because our method with alignment can detect fainter emission, originally embedded in the noise, due to the effect of de-correlation.  
Furthermore, our method gives a better constraint on the inner and outer radii of the N$_2$D$^+$ (3--2) ring, where emission is only marginally detected with method (1) and (2).
In conclusion, our method of alignment can measure intensity profiles of weak molecular lines at high significance, providing constraints on size, width, and depth of cavities, gaps, and rings (if present) in protoplanetary disks where molecular-line intensity is low.

We further compute the expected S/N enhancement (Equation \ref{Rsn}) to compare with our observational results (Figure~\ref{Ir}b). 
We adopte $\Delta V_{\rm int} \sim 0.3 \times (r/{\rm 100\ AU})^{-0.3}$.
For the S/N enhancement compared to method (1),
we substitute $\Delta V_{\rm na}$ (which is a function of radius) with a constant velocity width of 9.3 km s$^{-1}$,
that is the velocity width we integrated to generate the moment 0 maps. 
The S/N enhancements in DCO$^+$ (3--2), DCN (3--2), N$_2$D$^+$ (3--2), and H$_2$CO (3$_{0,3}$--2$_{0,2}$) are consistent with expectation, except at radii smaller than the synthesized beam size where the observed enhancement is lower, 
and Equation \ref{Rsn} is not valid any longer.

\begin{figure*}
\epsscale{1}
\plotone{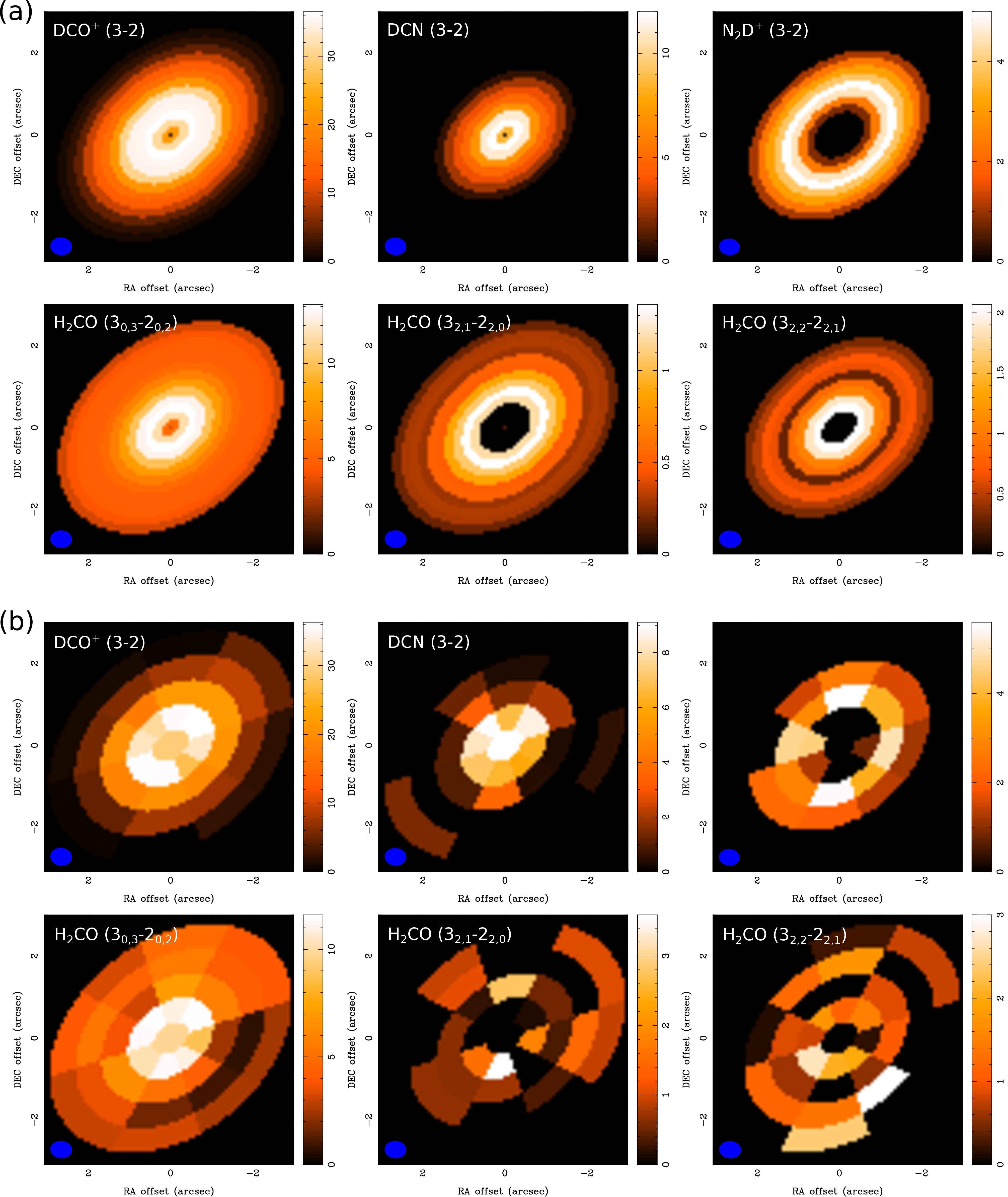}
\caption{Reconstructed moment 0 maps of DCO$^+$ (3--2), DCN (3--2), N$_2$D$^+$ (3--2), and H$_2$CO (3$_{0,3}$--2$_{0,2}$), (3$_{2,2}$--2$_{2,1}$), and (3$_{2,1}$--2$_{2,0}$) in HD~163296 from intensity profiles from our method of azimuthally averaging (a) over 2$\pi$ and (b) over azimuthally distinct sectors with alignment.}
\label{Irmap}
\end{figure*}

For a visual impression, we reconstruct moment 0 maps from the radial profiles (Figure~\ref{Irmap}a), noting that the actual emission distributions appear not to be fully axisymmetric (Figure~\ref{mom0}). 
The intensity distributions of H$_2$CO (3$_{2,2}$--2$_{2,1}$) and (3$_{2,1}$--2$_{2,0}$) are similar to those of DCO$^+$ (3--2) and H$_2$CO (3$_{0,3}$--2$_{0,2}$), and the intensities peak at radii around $\sim$100 AU. 
In addition, all three H$_2$CO transitions show flat distributions at radii beyond $\gtrsim$200 AU. This is different from DCO$^+$ (3--2) and DCN (3--2) that show more steeply declining profiles.
The ring-like structure in DCN (3--2) is possibly located closer (at a radius of $\sim$50 AU) to the center than those in DCO$^+$ and H$_2$CO, 
while the ring-like structure of N$_2$D$^+$ is further outside than all the other lines at around 200 AU.
This trend is consistent with the observational results in DCO$^+$ (4--3) and N$_2$H$^+$ (3--2) in HD~163296 (Mathews et al.~2013; Qi et al.~2015), which show N$_2$H$^+$ (3--2) having a larger inner cut-off radius than DCO$^+$ (4--3). 
Similarly, we can also measure mean integrated intensities of azimuthally distinct sectors to investigate asymmetric intensity distributions. 
We divide the outer disk at $r \gtrsim 0\farcs45$ (the beam size) into azimuthally and radially distinct sectors, 
each sector spanning 1.5 times the beam size along the radial axis and 45$\arcdeg$ in azimuth.  
The inner disk inside $r < 0\farcs45$ is still azimuthally averaged over 2$\pi$ because position angles of individual pixels cannot be computed accurately for such small radii with this resolution. 
For each sector, we align and stack spectra from every pixel within its radial and azimuthal range, divide the stacked spectra by the number of pixels included, and measure the total integrated intensity.
The reconstructed moment 0 maps are presented in Figure \ref{Irmap}b.
In the original moment 0 map (Figure~\ref{mom0}), DCO$^+$ (3--2) is brighter towards the northwest and the southeast, and N$_2$D$^+$ (3--2) shows brighter clumps towards north and south. 
These features are also identified in the reconstructed maps (Figure~\ref{Irmap}b). 
The original bright DCN (3--2) emission towards east is mostly in the innermost bin that is now averaged over 2$\pi$.
Hence, this feature is not clear in the reconstructed map (Figure~\ref{Irmap}b).
A fainter emission towards southwest is now revealed in H$_2$CO (3$_{0,3}$--2$_{0,2}$), and the previous 
non-detections now show patchy detections in H$_2$CO (3$_{2,2}$--2$_{2,1}$) and (3$_{2,1}$--2$_{2,0}$).
The detailed interpretation of the different distributions of these molecular lines is beyond the scope of the present paper and is not discussed here.

In summary, our results demonstrate that by aligning spectra with different centroid velocities from different positions in a disk and stacking them, 
we are able to enhance the S/N of molecular-line data
and materialize spectra of molecular lines that are originally undetectable.  
Our method can significantly lower the detection limit of molecular lines and, thus, be applied to search for faint molecular lines in protoplanetary disks.
Furthermore, because of the S/N enhancement, 
intensity profiles of molecular lines can be measured more accurately and in smaller bins, which is equivalent to achieving higher spatial resolutions.  
Molecular-line images can also be generated at higher angular resolutions because more weighting can be given to longer baselines, which 
typically are noisier, but their S/N can be enhanced with this method.
Finally, our method can be applied not only to molecular-line data of protoplanetary disks but to any systems having ordered kinematic patterns.

\acknowledgments
This paper makes use of the following ALMA data: ADS/JAO.ALMA\#2013.1.01268.S. ALMA is a partnership of ESO (representing its member states), NSF (USA) and NINS (Japan), together with NRC (Canada) and NSC and ASIAA (Taiwan), in cooperation with the Republic of Chile. The Joint ALMA Observatory is operated by ESO, AUI/NRAO and NAOJ. We thank all the ALMA staff supporting this work. 
P.M.K. acknowledges support from an Academia Sinica Career Development Award.
P.M.K, N.H., C.-F.L. and S.T acknowledge support from the Ministry of Science and Technology (MOST) of Taiwan
through grants MOST 104-2119-M-001-019-MY3, MOST 104-2119-M-001-016, MOST 104-2119-M-001-015-MY3
and MOST 102-2119-M-001-012-MY3.



\begin{thebibliography}{}
\bibitem[ALMA Partnership et al.(2015)]{ALMA15} ALMA Partnership, Brogan, C.~L., P{\'e}rez, L.~M., et al.\ 2015, \apjl, 808, L3 
\bibitem[Cald{\'u}-Primo et al.(2013)]{2013AJ....146..150C} Cald{\'u}-Primo, A., Schruba, A., Walter, F., et al.\ 2013, \aj, 146, 150 
\bibitem[de Gregorio-Monsalvo et al.(2013)]{Gregorio13} de Gregorio-Monsalvo, I., M{\'e}nard, F., Dent, W., et al.\ 2013, \aap, 557, A133 
\bibitem[Delhaize et al.(2013)]{Delhaize13} Delhaize, J., Meyer,M.~J., Staveley-Smith, L., \& Boyle, B.~J.\ 2013, \mnras, 433, 1398 
\bibitem[Dutrey et al.(2007)]{Dutrey07} Dutrey, A., Henning, T., Guilloteau, S., et al.\ 2007, \aap, 464, 615
\bibitem[Dutrey et al.(2014)]{Dutrey14} Dutrey, A., Semenov, D., Chapillon, E., et al.\ 2014, in Protostars and Planets VI, ed., H. Beuther, R. S. Klessen, C. P. Dullemond, and T. Henning (Tucson, AZ: Univ. Arizona Press), 317 
\bibitem[Higdon et al.(2015)]{2015ApJ...814L...1H} Higdon, J.~L., Higdon, S.~J.~U., Mart{\'{\i}}n Ruiz, S., \& Rand, R.~J.\ 2015, \apjl, 814, L1 
\bibitem[Hughes et al.(2008)]{Hughes08} Hughes, A.~M., Wilner, D.~J., Qi, C., \& Hogerheijde, M.~R.\ 2008, \apj, 678, 1119 
\bibitem[Isella et al.(2007)]{Isella07} Isella, A., Testi, L., Natta, A., et al.\ 2007, \aap, 469, 213 
\bibitem[Mannings\& Sargent(1997)]{Mannings97} Mannings, V., \& Sargent, A.~I.\ 1997, \apj, 490, 792 
\bibitem[Mathews et al.(2013)]{Mathews13} Mathews, G.~S., Klaassen, P.~D., Juh{\'a}sz, A., et al.\ 2013, \aap, 557, A132 
\bibitem[Pi{\'e}tu et al.(2007)]{2007A&A...467..163P} Pi{\'e}tu, V., Dutrey, A., \& Guilloteau, S.\ 2007, \aap, 467, 163 
\bibitem[Qi et al.(2011)]{Qi11} Qi, C., D'Alessio, P., {\"O}berg, K.~I., et al.\ 2011, \apj, 740, 84 
\bibitem[Qi et al.(2015)]{Qi15} Qi, C., {\"O}berg, K.~I., Andrews, S.~M., et al.\ 2015, \apj, 813, 128 
\bibitem[Qi et al.(2013)]{2013ApJ...765...34Q} Qi, C., {\"O}berg, K.~I., \& Wilner, D.~J.\ 2013, \apj, 765, 34 
\bibitem[Rosenfeld et al.(2013)]{Rosenfeld13} Rosenfeld, K.~A., Andrews, S.~M., Hughes, A.~M., Wilner, D.~J.,  \& Qi, C.\ 2013, \apj, 774, 16
\bibitem[van den Ancker et al.(1998)]{Ancker98} van den Ancker, M.~E., de Winter, D., \& Tjin A Djie, H.~R.~E.\ 1998, \aap, 330, 145 

\end{thebibliography}
\end{document}